\documentclass[12pt, a4paper]{article}
\usepackage{bm}
\usepackage{graphicx}
\newcommand{\sign}{\mathop{\mathrm{sign}}\nolimits}
\title{Conductance through the disclination dipole defect in metallic carbon nanotubes}

\author{D.V. Kolesnikov$^1$ and V.A. Osipov$^2$\\
Bogoliubov Laboratory of Theoretical Physics,\\ Joint
Institute for Nuclear Research, 141980 Dubna,\\ Moscow region,
Russia\\
e-mail: $^1$kolesnik@theor.jinr.ru, $^2$osipov@theor.jinr.ru}

\begin{document}
\maketitle \abstract{ The electronic transport properties of a
metallic carbon nanotube with the five-seven disclination pair
characterized by a lattice distortion vector are investigated. The
influence of the disclination dipole includes induced curvature
and mixing of two sublattices. Both these factors are taken into
account via a self-consistent perturbation approach. The
conductance and the Fano factor are calculated within the
transfer-matrix technique.\\

 }
\section*{INTRODUCTION}\label{intro}
Transport properties of variously shaped carbon nanostructures and
graphene are of great practical and theoretical interest. In
particular, the conductivity of carbon nanotubes is currently the
subject of wide investigations. Actually, there are many types of
defects  (such as vacancies and vacancy pairs \cite{vac0}, adatoms
\cite{adatom}, structural defects, etc.) which influence the
conductivity of carbon nanotubes. For instance, the electronic and
transport properties of nanotubes containing vacancy pairs were
found to be extremely sensitive to the sublattice positioning of the
pair \cite{vac1}. Moreover, it was observed that they substantially
affect the metallic or semiconducting character of the tube. The
structural (topological) defects in carbon nanotubes are mainly
five-seven disclination pairs~\cite{57} and the Stone-Wales defects
("quadrupoles", 5-8-5 or 5-7-7-5 defects \cite{sw}). The Stone-Wales
structural defects were found to influence at least the local
electronic properties of the nanotube. It is also known that
disclinaton topological defects (fivefolds) can convert planar
graphene surface into the conical one with a marked modification of
the electronic states \cite{cone}. In the nanotubes, however, the
presence of an isolated pentagonal ring is not allowed and the
simplest topological defect is the 5-7 disclination dipole (DD).
The simple model for the junction between two metallic tubes was
investigated in~\cite{tamura_tsukada}.

In this paper, we study the electronic transport properties of a
metallic nanotube containing closely spaced 5-7 disclination
dipole (i.e., the junction composed of two semi-infinite $metallic$ nanotubes).
The paper is organized as follows. In Section \ref{s1},
the geometrical characteristics of the nanotube with the
DD-induced curvature are discussed. The structure of a junction
consisting of two nanotubes with different diameters is presented.
In order to design the shape of the tube we introduce two
parameters: the distortion vector $\vec b$ which influences both
the radius and the chirality of the tube, and the phenomenological
parameter $\xi_0$ characterizing the size of the curved region due
to elastic relaxation. In Section \ref{s2}, the effective Dirac
equation is formulated, and a standard transfer matrix method is
used to calculate the conductance and the Fano factor of the
structure. The obtained results are discussed in Section
\ref{concl1}.
\section{GEOMETRY}\label{s1}
A metallic or semiconducting character of the tube is governed by
the translational vector $\vec T_0(n,m)$, where $(n,m)$ are the
numbers of steps along two unit cell vectors in the honeycomb
lattice. The boundary conditions for the tube with a translational
vector $\vec T$ transforms to the angular vector field $a_\varphi$
which dependis on  chirality as $(n+m)\; mod\; 3$. This field,
along with the angular momentum, subsequently generates the
mass-type term in the effective Dirac equation. The value of the
produced gap in the energy spectrum is determined as
$2(j-a_\varphi)\hbar V_F/R_0$, where $j$ is the angular momentum
and $R_0$ is the tube radius. When this term vanishes, the
metallization of the nanotube occurs for $j=0$ (see \cite{melet}
for detail).

As for the fivefold-sevenfold pair in the nanotube, it appears to
be the source of translational-type holonomy \cite{gfields},
therefore it should produce additional gauge vortex field.
However, in this paper we restrict our consideration to a special
case when the condition $(n+m)\; mod \; 3=0$ is fulfilled at
$both$ sides. In other words, both tubes far from the junction
region are suggested to be metallic. The gauge field produced by
the dipole source should affect the chirality but the value
$(n+m)\; mod\; 3$ remains untouched. Additionally, at low energies
below the first gap value ($|E|<<h V_F/R_0$, where $V_F$ is the
Fermi velocity and $R$ is the tube radius) one can take into
account for the estimation of first-order perturbation only the
main conducting channel corresponding to the lowest angular
momentum.

As is known, the dislocation-type defect touches both the
chirality and the radius of the carbon nanotube. In the general
case, the tube with DD can be characterized by the distortion
vector $\vec b$: for the translational vector $\vec T$ one can
find, that it has the value $\vec T_0,\;T_0=2\pi R_0$ on the one
side and $\vec T_0+\vec b$ on the other.

\begin{figure}
\resizebox{0.55\columnwidth}{!}{%
  \includegraphics{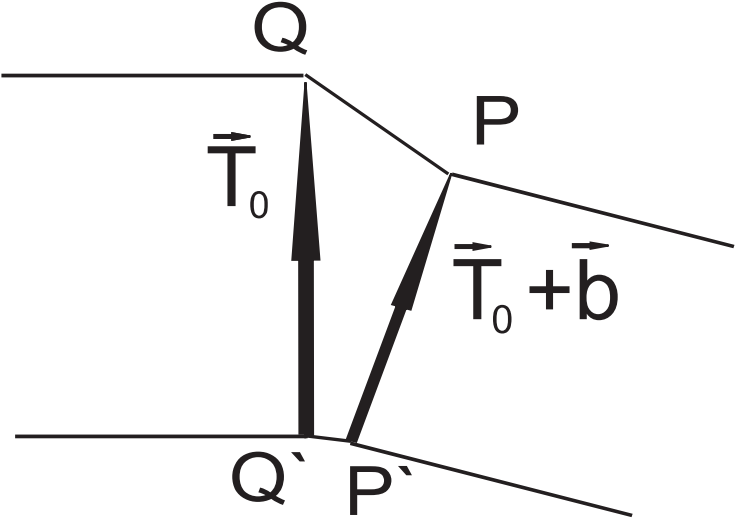}
}
\caption{The schematic structure of the nanotube in the defect
region. The fivefold is located at the point Q (Q'), and the
sevenfold - at the point P (P'). Both the length and the direction
of the translational vector $\vec T_0$ are changed in the presence
of the 5-7 disclination dipole.}
\label{fig1l}
\end{figure}

Trying to describe a shape of the nanotube, it is useful to
perform a development of the junction region onto a 2D plane (see
Fig. \ref{fig1l}).

Generally, the DD with a distortion vector
$\vec b$ changes both the length and the direction of the
translational vector. As a result, on the right side the shape of
the tube is also a cylinder with the radius $|\vec T_0+\vec
b|/(2\pi)$, displaced by the angle between vectors $\vec T_0$ and
$\vec T_0+\vec b$. In this paper, we consider $b/T_0$ as a small
parameter. As an example, this could be the vector $\vec b(2,1)$
which preserves the metallicity of the tube. Reverting to the
shape of the tube we should roll up the development. This
resembles the "cut-and-glue" procedure for a disclination in the
nanocone \cite{cone}. This procedure determines an explicit relative
positions for the cylinders (both the angle and the distance
 between the axis lines) as a function of $\vec b$. For the
  exact shape of the structure in the junction region, let us construct a phenomenological approximation, which
satisfies all the boundary conditions described above.

\begin{figure*}
\resizebox{0.75\columnwidth}{!}{%
  \includegraphics{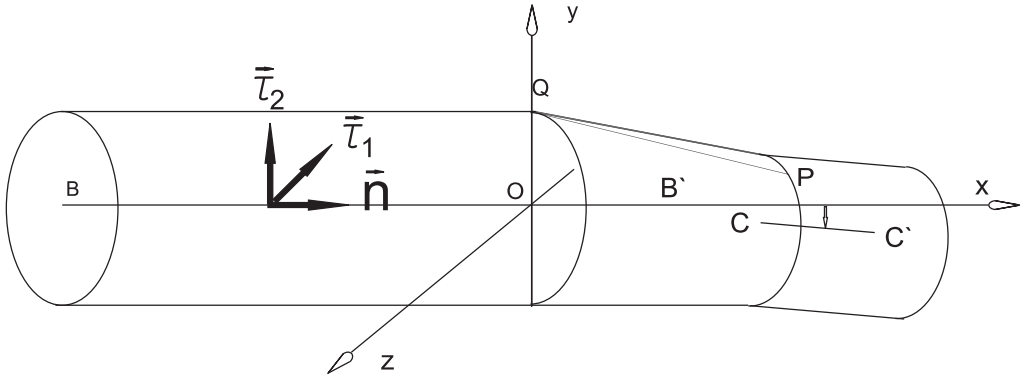}
}
\caption{A schematic picture of the nanotube surface. The fivefold
is located at the point Q, and the sevenfold - at the point P. The
thick tube axis BB' coincides with the x-axis. The thin tube axis
CC' is shifted due to the disclination dipole QP.}
\label{fig2l}
\end{figure*}

Let us associate the development with the xy-plane and choose the
axis along the tube on the left side as the x-axis. In this case,
one has $\vec T_0(0,T_0,0)$ and $\vec T(\xi)=\vec T(T_x,T_y,0)$.
It is convenient to introduce the new frame with the vector $\vec
n=(\vec T \times \vec e_z)/T$ ($\vec e_i\, (i=x,y,z)$ are the
orthogonal basis vectors) which follows the tube axis (see Fig.
\ref{fig2l}).
\begin{figure*}
\resizebox{0.75\columnwidth}{!}{%
  \includegraphics{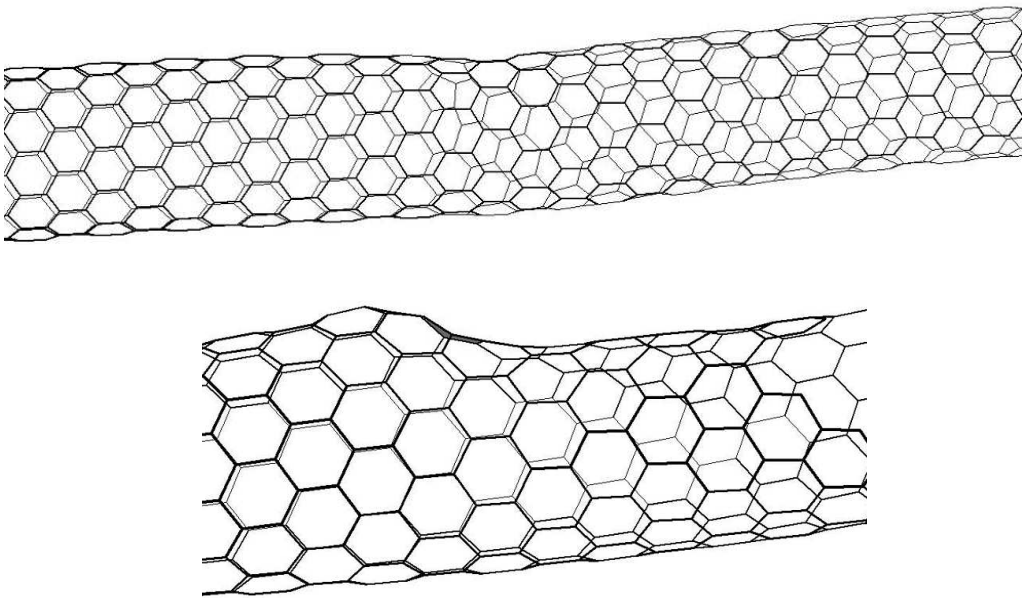}
}
\caption{A molecular-dynamics simulation for the shape of the
(12,0)-(10,1) junction \cite{gluxova}. Bottom: in the junction
region the shifting of the tube axis can be observed.}
\label{fig2al}
\end{figure*}

The orthogonal to $\vec n$ vectors  are $\vec \tau_1=-\vec e_z$ and
$\vec \tau_2=\vec \tau_1 \times \vec n=\vec T/T$. To take into
account the disclination dipole, which is situated at the top of the
cylinder, one needs to describe the skewed conical surface between
$Q$ and $P$ points. Actually, the DD has two effects: it changes the
radius and shifts the surface of the tube in the direction normal to
the tube axis. Since the radius $R(\xi)=T(\xi)/(2\pi)$ changes its
value on $\Delta R$, the ring will be shifted in the direction of
$\vec \tau_2$ by $\Delta R$, so that we must add the shift $\vec
\tau_2 T/(2\pi)$.

Finally, the surface is parametrized as following:
\begin{eqnarray}\label{rxi}
  \vec r=\vec n \xi+\vec \tau_2 \frac{T}{2\pi}+\vec \tau_1
  \frac{T}{2\pi}\cos\varphi+\vec \tau_2 \frac{T}{2\pi}\sin\varphi,
\end{eqnarray}
where $-\infty<\xi<\infty$ and $0\leq\varphi<2\pi$ are the normal
and the transversal coordinates, respectively.

The form of the tube with 5-7 rings is supposed to be
\begin{eqnarray}\label{r} \vec r (\xi
\frac{T_y}{T}+\frac{T_x}{2\pi}(1+\sin\varphi),-\xi
\frac{T_x}{T}
+\frac{T_y}{2\pi}(1+\sin\varphi),-\frac{T}{2\pi}\cos\varphi)
\end{eqnarray}
where we assume
\begin{equation}\label{Txi}
\vec T(\xi)=\vec T_0 + \frac{\vec b}{2}(\tanh \frac{\xi}{\xi_0}+1)
\end{equation}
to be the tube translation vector, depending on the coordinate
$\xi$, the distortion vector $\vec b$ and an effective parameter
$\xi_0$, which corresponds to the half-width of the surface region
curved by the DD. Since the surface of the nanotube has high but
finite Young modulus, one can expect that $\xi_0$ is of the same
order as $R_0$. Let us calculate the geometrical properties of the
tube in the first order in $\vec b$.

The metric tensor is found to be
\begin{eqnarray}\label{gmunu}
  g_{\xi\xi}=1+\xi\frac{(T^2)'}{T^2}-2\xi\frac{T'}{T}
  +2\frac{T_y T_x'-T_x T_y'}{2\pi
  T}(1+\sin\varphi)+{\cal O}(b^2)=1+\gamma_{\xi\xi},\nonumber \\
  g_{\xi\varphi}=\frac{1}{2}\left[\frac{(T^2)'}{4\pi^2}-2\xi\frac{T_y T_x '-T_x T_y '}{2\pi
  T}\right ]\cos\varphi
  +{\cal O}(b^2)=\gamma_{\xi\varphi},\nonumber\\
  g_{\varphi\varphi}=R_0^2+\frac{\vec T_0 \cdot\vec
  b}{4\pi^2}(\tanh\frac{\xi}{\xi_0}+1)
  +{\cal O}(b^2)=R_0^2+\gamma_{\varphi\varphi},
\end{eqnarray}
and
\begin{eqnarray}\label{gup}
  g^{\xi\xi}=1-\gamma_{\xi\xi},\;
  g^{\xi\varphi}=-\frac{\gamma_{\xi\varphi}}{R_0^2},\;
  g^{\varphi\varphi}=\frac{1}{R_0^2}-\frac{\gamma_{\varphi\varphi}}{R_0^4}.
\end{eqnarray}
The metrical connection
\begin{eqnarray}\label{Gdef}
  \Gamma^k_{\mu\lambda}=\frac{1}{2}g^{kl}(\partial_\mu g_{l\lambda}+\partial_\lambda g_{\mu l}-\partial_l g_{\mu \lambda})
\end{eqnarray}
is found to be
\begin{eqnarray}\label{Gs}
  \Gamma^\xi_{\xi\xi}=\partial_\xi \gamma_{\xi\xi}/2;\;
  \Gamma^\xi_{\xi\varphi}=\Gamma^\xi_{\varphi\xi}=\partial_\varphi
  \gamma_{\xi\xi}/2;\nonumber\\
  \Gamma^\xi_{\varphi\varphi}=\partial_\varphi\gamma_{\xi\varphi}-\partial_\xi\gamma_{\varphi\varphi}/2;\nonumber\\
  \Gamma^\varphi_{\xi\xi}=\partial_\xi\gamma_{\xi\varphi}/R_0^2-\partial_\varphi\gamma_{\xi\xi}/(2 R_0^2);\nonumber\\
  \Gamma^\varphi_{\xi\varphi}=\Gamma^\varphi_{\varphi\xi}=\partial_\xi\gamma_{\varphi\varphi}/(2
  R_0^2),
\end{eqnarray}and $ \Gamma^\varphi_{\varphi\varphi}=0$.
Tetradic coefficients $e^i_{\; \mu}$, which are determined by the
equation
\begin{eqnarray}\label{edef}
  g_{\mu\nu}=e^i_{\; \mu} e^k_{\; \nu} \delta_{ik},
\end{eqnarray}
are chosen to be
\begin{eqnarray}\label{etetra}
  e^1_{\; \xi}=1+\frac{\gamma_{\xi\xi}}{2};\,
  e^1_{\;\varphi}=\gamma_{\xi\varphi};\,
  e^2_{\; \xi}=0;\, e^2_{\; \varphi}=R_0+\frac{\gamma_{\varphi\varphi}}{2R_0}
\end{eqnarray}
and
\begin{eqnarray}\label{etetralow}
  e_1^{\; \xi}=1-\frac{\gamma_{\xi\xi}}{2};\,
  e_1^{\; \varphi}=0;\nonumber\\
  e_2^{\; \xi}=-\frac{\gamma_{\xi\varphi}}{R_0};\; e_2^{\;
  \varphi}=\frac{1}{R_0}-\frac{\gamma_{\varphi\varphi}}{2R_0^3}.
\end{eqnarray}
The spin connection coefficients $$(\omega_\mu)^{ab}=e^a_{\;
\nu}D_\mu e^{b\nu},$$ where $D_\mu:=\partial_\mu+\Gamma_\mu$ is a
covariant derivative, reads
\begin{eqnarray}\label{omegas}
  \omega_\xi^{\; 12}=-\frac{\partial_\xi
  \gamma_{\xi\varphi}}{R_0}+\frac{\partial_\varphi
  \gamma_{\xi\xi}}{2R_0}
  =\left [-\frac{(T^2)''}{8\pi^2 R_0}+\xi \frac{T_y T_x''-T_x T_y''}{4\pi^2 R_0^2} \right
  ]\cos\varphi;\nonumber\\
  \omega_\varphi^{\; 12}=-\frac{\partial_\xi
  \gamma_{\varphi\varphi}}{2 R_0}=-\frac{(T^2)'}{8\pi^2 R_0};\;
  \omega_\mu^{\;21}=-  \omega_\mu^{\;12}.
\end{eqnarray}
\section{CONDUCTIVITY AND THE FANO FACTOR}\label{s2}
From now on we set $E_F=0,\, V_F=\hbar=1$. The Dirac equation on
the curved surface is defined as
\begin{eqnarray}\label{dirac}
  -i\sigma_i e_i^{\; \mu}(\nabla_\mu-iA_\mu)\psi=E \psi,
\end{eqnarray}
with $\sigma_i$ being the Pauli matrices ($i={1,2}$), $\psi$ is a
2-component spinor wavefunction, and $A_\mu\; (\mu={\xi,\,
\varphi})$ being the DD field. The derivative includes a spin
connection term $\Omega_\mu$ and it is written as
$$\nabla_\mu=\partial_\mu+\Omega_\mu,\;
\Omega_\mu=\frac{1}{8}\omega_\mu^{\; ab}[\sigma_a,\sigma_b].$$ In
the linear in $\vec b$ approximation the Dirac equation takes the
form
\begin{eqnarray}\label{diracd}
  -i\sigma_1
  e^{1\xi}(\nabla_\xi-iA_\xi)\psi
  -i\sigma_2(e^{2\xi}\partial_\xi+e^{2\varphi}(\nabla_\varphi-iA_\varphi))\psi=E\psi,
\end{eqnarray}
with the covariant derivatives written as
\begin{eqnarray}\label{Omegas}
  \nabla_\varphi=\partial_\varphi-\frac{(T^2)'}{16 \pi^2
  R_0}\sigma_3;\nonumber\\
\nabla_\xi=\partial_\xi+[-\frac{(T^2)''}{8 \pi^2
  R_0}+\xi\frac{T_y T_x''-T_x T_y''}{4\pi^2 R_0^2}]\frac{\sigma_3}{2}\cos\varphi.
\end{eqnarray}
Substituting (\ref{Omegas}) into (\ref{diracd}) we find both the
undisturbed Dirac operator $\hat{\cal D}_0$
\begin{eqnarray}\label{d0}
  \hat{\cal
  D}_0=-i\sigma_1\partial_\xi-i\frac{\sigma_2}{R_0}\partial_\varphi,
\end{eqnarray} and the perturbation
operator $\hat {\cal V}$
\begin{eqnarray}\label{v}
  \hat{\cal V}=i\sigma_1
  \frac{\gamma_{\xi\xi}}{2}\partial_\xi-i\sigma_1\Omega_\xi-\sigma_1A_\xi+
  i\sigma_2\frac{\gamma_{\xi\varphi}}{R_0}\partial_\xi
  -i\frac{\sigma_2}{R_0}\Omega_\varphi-\frac{\sigma_2}{R_0}A_\varphi+
  i\sigma_2\frac{\gamma_{\varphi\varphi}}{2R_0^3}\partial_\varphi.
\end{eqnarray}

It should be mentioned that the perturbation~(\ref{v}) includes
the last term, which is not localized (differs from zero value at
$\xi\rightarrow\infty$). Therefore an analysis of (\ref{d0})
and~(\ref{v}) differs from a typical scattering problem. To
simplify the problem, let us restrict our consideration to the
lowest term with the zero angular momentum, that is only a plain
wave $\psi(\xi,\varphi)=(1,\pm 1)^T e^{ik\xi}$ is considered. In
this case in (\ref{d0}) and (\ref{v}) only the zero-momentum terms
should be taken into account in the Fourier series for the angle-dependent
part of the wavefunction.

As a result, instead of~(\ref{d0}) and~(\ref{v}) there should apper an equation
for the zero-momentum part of the wavefunction with the Dirac hamiltonian
 $\hat{\cal D}_0=-i\sigma_1\partial_\xi$ and the perturbation
\begin{eqnarray}\label{vf}
  \hat{\cal V}=i\sigma_1\frac{(\vec b\times\vec{T}_0)_z}{8\pi^2 R_0 \xi_0\cosh^2(\xi/ \xi_0)}\partial_\xi-\sigma_1<A_\xi>_\varphi
  -\sigma_1\frac{\vec b\cdot\vec{T}_0}{16\pi^2 R_0^2 \xi_0\cosh^2(\xi/
  \xi_0)}-\frac{\sigma_2}{R_0}<A_\varphi>_\varphi.
\end{eqnarray}
Here $<>_\varphi$ denotes the averaging over $\varphi$ and $(\vec b
\times \vec T_0)_z=b_x T_{0y}-T_{0x} b_y$. Now the
perturbation~(\ref{vf}) is localized near $\xi=0$ with a half-width
$\xi_0$. In order to remove the differentiation in (\ref{vf}) one
can multiply the operator $\hat{\cal D}_0+\hat{\cal V}$ by
$$\left (1+\frac{(\vec b\times\vec{T}_0)_z}{8\pi^2 R_0 \xi_0\cosh^2(\xi/
\xi_0)}\right ).$$ In this case, the additional potential $$I
E(\vec b\times\vec{T}_0)_z/(8\pi^2 R_0 \xi_0\cosh^2(\xi/ \xi_0))$$
appears, where $I$ is a unit matrix.

The field $\vec A$ describes the DD-induced valley mixing. It may
be written as $A_j=\mu\epsilon_{jk}\partial_k G$ where $j,k=\xi,\,
\varphi$, $\mu$ is an effective "charge" of each disclination, $G$
is a dipole potential, which satisfies the equation
$$
\nabla^2 G(\xi,\varphi)=\frac{2\pi}{R} [d_\xi
\delta'(\xi)\delta(\varphi)+d_\varphi\delta(\xi)\delta'(\varphi)/R_0],
$$
$\nabla^2=\partial^2_\xi+\partial^2_\varphi/R_0^2$,
$\epsilon_{jk}$ is a unit antisymmetric tensor, $\delta(x)$ and
$\delta'(x)$ are the Dirac delta function and it's derivative, and
$\vec d$ is a vector connecting five- and sevenfold (it has the
same order as $\vec b$ \cite{defgeom}). Explicitly, one can find
$G$ in a form
\begin{equation}\label{dvec}
G(\xi,\varphi)=-\frac{2\pi^2}{R_0}\sum_n [d_\xi \sign(\xi)+\frac{i
d_\varphi}{R}\sign(n)] e^{in\varphi}e^{-|n\xi|/R_0}.
\end{equation}
For the zero momentum $n=0$, the field $\vec A$ has the form
\begin{eqnarray}\label{wf}
A_\xi=0,\; A_\varphi=\frac{4\pi^2 \mu b_\xi}{R_0}\delta(\xi),
\end{eqnarray}
where $\delta(\xi)$ is a Dirac delta function. Finally, the
perturbation operator looks as following:
\begin{eqnarray}\label{vf2}
  \hat{\cal V}=-I\frac{(\vec b\times\vec{T}_0)_z\, E}{8\pi^2 R_0 \xi_0\cosh^2(\xi/ \xi_0)}
  -\sigma_1\frac{\vec b\cdot\vec{T}_0}{16\pi^2 R_0^2 \xi_0\cosh^2(\xi/ \xi_0)}
  -\sigma_2\frac{4\pi^2\mu d_\xi}{R_0^2}\delta(\xi).
\end{eqnarray}

Notice that the last term in (\ref{vf2}) corresponds to the
$\xi$-dependent mass term $\sigma_2 m(\xi)$ in the Dirac equation.
Let us calculate the transfer matrix and the conductance according
to the method described in \cite{titov} and~\cite{ballistic}. To transform the
initial Dirac equation to the diagonal form, the unitary rotation
is performed: $\psi\rightarrow {\cal{L}} \psi,\,\hat{\cal
V}\rightarrow{\cal{L}}\hat{\cal V}{\cal{L}^\dag},\,
\sigma_i\rightarrow{\cal{L}}\sigma_i{\cal{L}^\dag},\; {\cal{L}}
=(\sigma_1+\sigma_3)/\sqrt{2}.$ For zero angular momentum, the
equation for the transfer matrix reads
\begin{eqnarray}\label{teq}
  {\cal T} ( \frac{L}{2},-\frac{L}{2} )={\cal T}^{(0)} ( \frac{L}{2},-\frac{L}{2} )
  -i\int_{-\frac{L}{2}}^{\frac{L}{2}}
  dx {\cal T}^{(0)}(\frac{L}{2},x ) \sigma_3 \hat{\cal V}(x) {\cal T} (x,-\frac{L}{2} ) ,
\end{eqnarray}
where ${\cal T}^{(0)}(x_2,x_1)=\exp(i\sigma_3 E(x_2-x_1))$ and $L$
is the length of the tube. The perturbation (\ref{vf2}) takes the
form
\begin{eqnarray}\label{pert}
  \sigma_3 \hat{\cal{V}}=-\sigma_3\frac{(\vec b\times\vec{T}_0)_z \, E}{8\pi^2 R_0 \xi_0\cosh^2(\xi/
  \xi_0)} +I\frac{\mu b_y}{\xi^2+\pi^2 R_0^2}
  -\nonumber\\
  -I\frac{\vec b\cdot\vec{T}_0}{16\pi^2 R_0^2 \xi_0\cosh^2(\xi/
  \xi_0)}-i\sigma_1\frac{4\pi^2 \mu d_\xi}{R_0^2}\delta(\xi).
\end{eqnarray}

\begin{figure*}
\resizebox{0.65\columnwidth}{!}{%
  \includegraphics{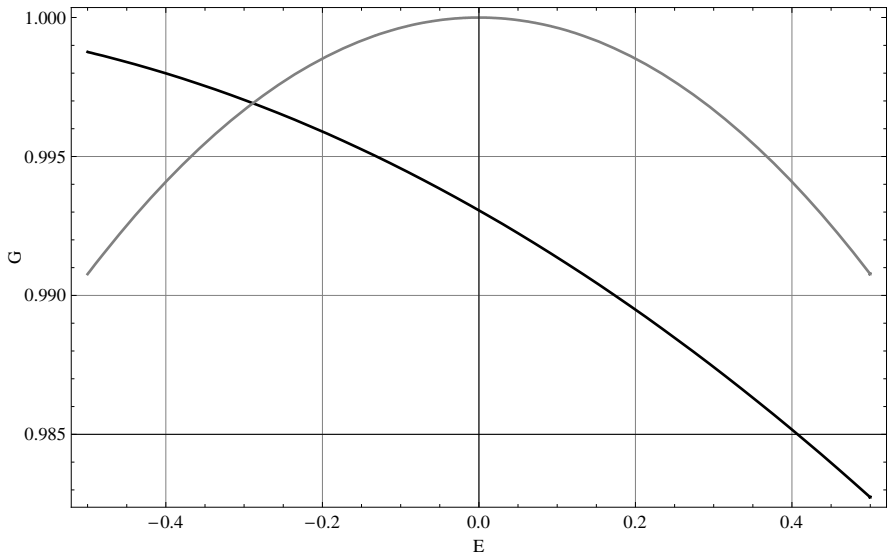}
}
\caption{Conductance $G$, in the units of $h/(4e^2)$, depending on
the energy $E$ in the units of $hV_F/R_0$, through the metallic
junction (9,0) - (11,1) tubes (black line) and (9,0) - (10,2)
(gray line).}
\label{fig3}       
\end{figure*}

We assume that $L$ is much larger than $\xi_0$.  In the linear in
$\vec b$ approximation one can replace ${\cal T}$ by ${\cal
T}^{(0)}$ in the integral in (\ref{teq}).  In our consideration,
the direct intervalley scattering is not taken into account. Thus
one can consider two inequivalent $K$-points (valleys) as two
different channels coupled by the non-Abelian field. Therefore,
one should perform a replacement $\mu\rightarrow\mu\tau_2$ with
$\mu=1/4$ and $\tau_2$ being the Pauli matrix in the channel space
representing the mixing of sublattices due to pentagonal and
hexagonal defects (see \cite{cone} for detail). By substituting
(\ref{pert}) into (\ref{teq}) and taking the limits of integration
to be infinity, one obtains the transfer matrix in the form
\begin{eqnarray}\label{tpert-na}
  {\cal T}=\exp(i\sigma_3 E L)+i  [
  \frac{\vec b\cdot\vec T_0}{8\pi^2 R_0^2}
  +\sigma_3 \frac{(\vec b\times\vec T_0)_z E}{4\pi^2
  R_0} ] \exp(i\sigma_3 E L)
  +\mu
  d_\xi \sigma_1\tau_2\frac{4\pi^2}{R_0^2}.
\end{eqnarray}
The scattering matrix $(t^{\dag})^{-1}$ is defined by the
upper-left element of the transfer matrix, so that the last term
in (\ref{tpert-na}) is neglected. The conductance is defined as
$G=(4e^2/h)Tr[t^\dag t]$, where the trace is taken over the
channel space. It takes the form
\begin{eqnarray}\label{Gcond-na}
  G=\frac{4e^2}{h}[1-\left( \frac{\vec b\cdot\vec T_0}{2 T_0^2}
  +\frac{(\vec b\times\vec T_0)_z \epsilon}{T_0^2}
  \right ) ^2],
\end{eqnarray}
where $\epsilon=E R_0$ is a dimensionless energy (in the absolute
units of $hV_F/R_0$). Correspondingly, the Fano factor
$F=1-Tr[(t^\dag t)^2]/Tr[t^\dag t]$ reads
\begin{eqnarray}\label{fano-na}
   F={\left ( \frac{\vec b\cdot\vec T_0}{2 T_0^2}+\frac{(\vec b\times\vec T_0)_z \epsilon}{T_0^2}
   \right )}^2.
\end{eqnarray}
One can see that the mass-type term (the DD term) does not
influence both the conductance and the Fano factor since only
linear terms in $\vec b$ are taken into account (the mass is
expected to be the second-order term). Depending on the angle between
$\vec T_0$ and $\vec T_0+\vec b$, two types of energy-dependent behaviour appear.
For the angle less than $\pi/2$, both the scalar and "vector" products
in (\ref{Gcond-na}) and (\ref{fano-na}) are of the same sign. In
this case, the linear in energy term in (\ref{Gcond-na})
leads to a decrease of the conductance with increasing energy (and
vice versa), see Fig.\ref{fig3} (black line). This case corresponds
to slightly curved tubular structures. For the second type of behaviour,
when the angle is greater than $\pi/2$ the increasing of conductance is expected with energy increasing. A special case is presented on Fig.\ref{fig3} (gray line), with $\vec T_0$ and $\vec b$ orthogonal, where a transmission coefficient is equal to unity at the Fermi energy $\epsilon=0$. However, in this case the higher-order perturbation terms should be included.

\section*{CONCLUSION}\label{concl1}
In this paper, we have investigated the electronic transport
properties of metallic nanotubes with the disclination dipole
defect. The case of disclination dipoles preserving the chirality
(metallic character of the tube) was considered. This assumption
simplifies the problem under consideration and allows us to
construct a self-consistent expansion scheme for the effective
Dirac equation. The perturbation includes both the curvature term,
which appear as the Gaussian-like smooth localized potential, and the
DD-induced term. The standard transfer-matrix approach is used to
calculate the conductance and the Fano factor of the structure.

As the marked effect of the curvature, the energy-dependent term
in both the conductance and the Fano factor appears. It should be
stressed that the tube length $L$ and the effective half-width
$\xi_0$ do not enter the final results. This differs from the case
of ballistic transport in graphene with various types of disorder
studied in \cite{ballistic}, where both the conductance and the
Fano factor were found to depend on the sample length (or, more
precisely, on $EL$).

Notice that in \cite{ando}, \cite{tamura_tsukada} a similar model for two metallic
nanotubes of different radiuses connected by the conical interface
was constructed. The transmission probability was found to
decrease when the difference of tube radiuses (which is determined
by $b_y$ in our paper) increases, which is in agreement with the
$~1-\alpha b^2$ behaviour in (\ref{Gcond-na}). The origin of this behaviour
within our model is purely geometrical, while in~ \cite{ando}
the boundary conditions for the wavefunction incorporates both the "geometrical" factor and the mixing of the K-points within the junction. As for the energy
dependence, in \cite{tamura_tsukada} the conductance was found to decrease with energy increasing for (17,17)-(18,21) junction, and subsequently the
increasing of conductance was found for (23,8)-(16,22) junction.
This fact is in general agreement with (\ref{Gcond-na}), where
the character of energy behaviour depends on the angle between translational
(chiral) vector and the distortion vector (the difference of chiral vectors).
At higher energies close to band edges, a more complex behaviour of conductance was predicted, which is absent in our model. The source of this
difference is twofold. First, instead of the boundary conditions
on the tube-cone interface in \cite{ando} and\cite{tamura_tsukada} we consider the model with smooth shape of the structure, so that the resonant behaviour
of wavefunction in the junction region does not appear in our
case. Second, for simplicity terms with higher momenta were not
included in our model and, as a result, localized and resonant
states are absent even at low energies. A self-consistent
inclusion of higher momenta in our approach is an important open
problem.

The disclination dipole is taken into account via the non-Abelian
field, which leads to the mixing of valleys in the low-energy
behaviour. The influence of this field was found to be negligible
within the perturbational approach compared to the
influence of curvature, in contrast with the case of ordinary vortex topological field sources
\cite{kolos1}. A dominating process of valley-mixing was taken
into account (as it is estimated for the scattering in carbon
nanocones \cite{cone}) instead of the real intervalley scattering,
which is expected to be omitted. This results in the correction of the conductance and the Fano factor
in the highest order in $\vec b$ to have the same form as for the
single-channel approximation. One should also note, that within
our model the chirality-dependence of the potential was neglected,
because both the left and right ends of the tube with the
disclination dipole are of the metallic-type (despite the fact
that the radius and exact chirality indices are different). In the
general case, one should expect the influence of the defect on
transport properties to be sensitive on the tube chiralities, as
it was observed in \cite{vac1}. Another important question is the
finiteness of the free electron path in the real tube. As it is
easy to see, the phenomenological localization length of the
curved area does not influence the transport properties within the
constructed approximation. Due to this fact, an effective
long-range influence of topological defects could possibly lead to
the non-trivial interplay with the mean free path length in the
tubes with disclination dipoles.

 We would like to thank O.E.Gluhova for the results of
molecular-dynamics calculations.

 This work has been supported by the Russian Foundation for Basic
Research under grant No. 08-02-01027.


\begin{thebibliography}{99}
\bibitem{vac0} \textit{Y.Ma et al.} Magnetic properties of vacancies in graphene and single-walled carbon nanotubes// New J. Phys. 2004. V. 6. P. 68.
\bibitem{adatom} \textit{P.O.Lethinen et al.} Structure and magnetic properties of adatoms on carbon nanotubes// Phys. Rev. B. 2004. V. 69. P. 155422.
\bibitem{vac1} \textit{T.Nakanishi, M.Igami and T.Ando.} Conductance Quantization in the Presence of Huge and Short-Range Potential in Carbon Nanotubes// Physica E. 2000. V. 6. P. 872.
\bibitem{57} \textit{L. Chico et al.} Pure Carbon Nanoscale Devices: Nanotube Heterojunctions// Phys. Rev. Lett. 1996. V. 76. P. 971.
\bibitem{sw} \textit{N.Chandra, S. Namilae, and C. Shet.} Local elastic properties of carbon nanotubes in the presence of Stone-Wales defects// Physical Review B. 2004. V.  69. P. 094101.
\bibitem{gfields} \textit{M.A.H.Vozmediano, M.I.Katznelson, and F.Guinea.} Gauge fields in graphene// arXiv:1003.5179. 2010.
\bibitem{cone} \textit{P. Lammert and V. Crespi.} Graphene cones: Classification by fictitious flux and electronic properties// Phys.Rev.B. 2004. V. 69. P. 035406.
\bibitem{tamura_tsukada} \textit{R.Tamura and M.Tsukada.} Relation between transmission rates and the wave functions in carbon nanotube junctions
// Phys.Rev.B. 2000. V. 61. P. 8548.
\bibitem{gluxova} \textit{O.E.Gluxova}. Private message.
\bibitem{melet} \textit{C.L.Kane and E.J.Mele.} Size, Shape, and Low Energy Electronic Structure of Carbon Nanotubes// Phys. Rev. Lett. 1997. V. 78. P. 1932.
\bibitem{defgeom} For the planar case, $|\vec d|=|\vec b|$; see \textit{O.V.Yazyev and S.G.Louie.} arXiv:1004.203. 2010.
\bibitem{titov} \textit{M. Titov.} Impurity-assisted tunneling in graphene// Europhys. Lett. 2007. V. 79. P. 17004.
\bibitem{ballistic} \textit{A. Schuessler et al.} Analytic theory of ballistic transport in disordered graphene// Phys. Rev. B. 2009. V. 79. P. 075405.
\bibitem{ando} \textit{H.Matsumura and T.Ando.} Topological Effects on Conductance of Nanotubes// Mol. Cryst. Liq. Cryst. 2000. V. 340. P. 725.
\bibitem{kolos1} \textit{D. V. Kolesnikov and V. A. Osipov.} The continuum gauge field-theory model for low-energy electronic states of icosahedral fullerenes// Europhys. J. B. 2006. V. 49. P. 465.
\end{thebibliography}
\end{document}